\documentclass[11pt,twoside]{article}

\usepackage{graphicx}
\usepackage{csquotes}
\usepackage[utf8]{inputenc}
\usepackage[english]{babel}
\usepackage{amsmath,amsfonts,amssymb}
\usepackage{mathpazo} 
\usepackage[scaled]{helvet}
\usepackage[T1]{fontenc}
\usepackage{url}
\usepackage[colorlinks=true, allcolors=blue]{hyperref}
\usepackage{authblk}
\usepackage{graphicx,xcolor}
\usepackage{wrapfig}
\usepackage{colortbl}
\usepackage{booktabs}
\usepackage{algorithm}
\usepackage[noend]{algpseudocode}
\usepackage{changepage}
\usepackage[left=35mm,right=25mm,top=25mm,bottom=25mm,headheight=15pt,headsep=10pt,twoside]{geometry}%
\usepackage[labelfont={bf,sf},%
labelsep=period,%
figurename=Fig.,%
singlelinecheck=off,%
justification=RaggedRight]{caption}
\definecolor{color0}{RGB}{0,0,0} 
\definecolor{color1}{RGB}{59,90,198} 

\DeclareRobustCommand{\eqref}[1]{Eq.~\ref{#1}}
\DeclareRobustCommand{\figref}[1]{Fig.~\ref{#1}}

\title{Analysis of  90 days operation of the gyroscope GINGERINO}

\author[1]{J~Belfi}
\author[1,2]{N~Beverini}
\author[1,2]{G.~Carelli}
\author[1]{A.~Di~Virgilio}
\author[1,2]{U.~Giacomelli*}
\author[1,2]{E.~Maccioni}
\author[1,3]{A.~Simonelli}
\author[1,2]{F.~Stefani}
\author[1]{G.~Terreni}

\affil[1]{INFN Sezione di Pisa, Pisa, Italy}
\affil[2]{University of Pisa, Pisa, Italy}
\affil[3]{Ludwig-Maximilians-University, Munich, Germany}

\affil[*]{umberto.giacomelli@pi.infn.it}

\begin{document}

\maketitle

\begin{abstract}
	\noindent
	GINGERINO is a square ring-laser prototype, which has been built to investigate the level of noise inside the Gran Sasso underground laboratory.
	It is meant for fundamental physics, but it provides suitable data for geophysics and seismology. Since May 2017 it is continuously acquiring data.
	The analysis of the first $90$ days shows that the duty cycle is higher than $95\%$, and the quantum shot noise limit is of the order of $ 10^{-10} $(rad/s)/$\sqrt{\mathrm{Hz}}$.
	It is located in a seismically active area, and it recorded part of the of central Italy earthquakes.
	Its high sensitivity  in the frequency band of fraction of Hz makes it suitable for seismology studies.
	The main purpose of the present analysis is to investigate the long term response of the apparatus.
	Simple and fast routines to eliminate the disturbances coming from the laser have been developed.
	The Allan deviation of the raw data reaches $10^{-7}$ after about $10^6s$ of integration time, while the processed data shows an improvement of one order of magnitude.
	Disturbances at the daily time scale are present in the processed data and the expected signal induced by polar motion and solid Earth tide is covered by those disturbances.
\end{abstract}

\section{Introduction}            
Ring Laser Gyroscopes (RLG) \cite{Schreiber2013} are, at present, the most precise sensors of absolute angular velocity for an Earth based apparatus. 
They are based on the Sagnac effect \cite{Sagnac1914} arising from a rigidly rotating ring laser cavity.
They are essential in estimating rotation rates relative to the local inertial frame in many contexts ranging from inertial guidance to angular metrology, from geodesy to geophysics.
The Gross Ring \textquotedblleft G\textquotedblright at the Wettzell Geodetic Observatory has obtained a resolution on the Earth rotation rate of about $15\times 10^{-14}$ $ rad/s$ with 4 hours of integration time ($3 \times 10^{-9}$ in relative units) \cite{Tercjak2017,privSchreiber}.
Such an unprecedented sensitivity shows that this class of instruments is even suitable to probe the spatio-temporal structure of the local gravity field.
Earth rotation is precisely calculated by the International Earth Rotation Service (IERS) through Very Long Baseline Interferometers (VLBI)\footnote{\href{www.iers.org}{www.iers.org}}.
VLBI has demonstrated highly accurate and stable determinations of the universal time, in large part because its very precise observations of quasars provide access to a nearly inertial celestial reference frame.
However, this can be obtained only in post-analysis, after a cumbersome  procedure and very heavy computation.
On the contrary, RLG data are directly linked to the instantaneous Earth rotational motion that can be continuously monitored.
A good  agreement between VLBI and G has been achieved by measuring and comparing the low frequency annual Chandler-Wobble effect, a free oscillation of the Earth \cite{Schreiber2011}. 
G is a single square ring laser and cannot discriminate among tilts and rotation. 
An improvement in sensitivity and a full 3-dimensional ring laser system would allow to integrate efficiently the data produced by VLBI and also to measure the Lense-Thirring effect.
GINGER (Gyroscopes IN GEneral Relativity) will aim at measuring the gravito-magnetic (Lense-Thirring) effect of the rotating Earth by means of an array of high sensitivity and accuracy ring lasers \cite{Tartaglia2017, DiVirgilio2017}.
In the weak-field approximation of Einstein equation, the Sagnac frequency $f_S$, seen by a RLG located in a laboratory on the Earth surface, with co-latitude $\theta$,  and with the axis contained in the meridian plane at an angle $\psi$  with respect to the zenith, can be written as \cite{DiVirgilio2017}: 
\begin{equation}
f_S = S\Omega_E[\cos{(\theta +\psi)}-2\frac{GM_E}{c^2R_E}\sin{\theta}\sin{\psi}+\frac{GI_E}{c^2R_E^3}({2\cos{\theta}\cos{\psi}+\sin{\theta}\sin{\psi}})]
\label{GR}
\end{equation}
where $S$ is the geometric scale factor for the RLG, $G$ the Newton gravitational constant, $\Omega_E= 7.29\times10^{-5}$ $rad/s$ the Earth instantaneous angular rotation velocity, $M_E$ the Earth mass, $c$ is the light velocity in vacuum, $R_E$ the Earth mean radius, and $I_E$ the Earth moment of inertia.
The first term corresponds to the standard Sagnac signal.
The second one, known as geodetic or De Sitter precession, is produced by the motion of the laboratory in the
curved space-time around the Earth.
The third one, known as Lense-Thirring precession (LT), is produced by the rotating mass of the Earth and it is proportional to the Earth angular momentum \cite{Tartaglia2017,Bosi2011}.
The last two terms are both relativistic and smaller than the classical Sagnac effect by a factor of $10^{-8}\div10^{-9}$, that is of the order of magnitude of the ratio between $R_E$ and the Schwarzschild radius of the Earth  $2GM_E/c^2$.
These effects should be observed as a difference between the rotation rate observed by the array of RLGs in the rotating frame of the laboratory, and the length of the day determined in the "fixed stars" inertial frame by IERS. 
Registering a perturbation that amounts to 1 part in a billion of the Earth rotational rate, requires an unprecedented sensitivity of the apparatus. 
An array of at least three ring lasers would allow to measure the Earth angular velocity and, having at disposal the time series of the daily estimate of Earth rotation vector from the IERS Service, it would be possible to isolate the Lense-Thirring contribution.
Moreover, as it has been discussed in recent papers \cite{DiVirgilio2017}, the environmental noises pose a severe limitation, if the apparatus is located on the Earth surface.
An underground location, far from disturbances as hydrology, temperature and atmospheric pressure changes, is essential for this challenging experiment, and LNGS (Laboratori Nazionali del GranSasso, the underground Italian national institute of nuclear physics, INFN) offers a suitable location.
LNGS underground laboratory exhibits a very high natural thermal stability and, being deep underground, it is not affected by top soil disturbances.
In order to quantify the advantages of the underground location, a square ring laser apparatus called GINGERINO has been installed inside LNGS \cite{Belfi2017}.
This installation is a prototype on the path to GINGER and to RLG fundamental physics apparatus, but at the same time provides unique information for geophysics.
It has been already shown that the resolution of GINGERINO is of the order of $0.1 nrad/s$ with an integration time of 1 second \cite{Belfi2017}.
At present for Earth based apparatus, this level of sensitivity can be obtained by optical gyroscopes only. 
Compared to G, the sensitivity of GINGERINO is about a factor 50 worse, nevertheless the present paper shows an important achievement, since it is the first evidence that continuous data taking with high duty cycle and high sensitivity can be obtained with  ring lasers based on hetero-lithic mechanical structure, as GINGERINO and ROMY (ROtational Motion seismologY)\footnote{\href{www.romy-erc.eu}{www.romy-erc.eu}}. It is well accepted that future effort must be focused on the realisation of arrays of ring lasers, ROMY being the first example, and that the monolithic structure used for G cannot be expanded to form and array.
Any tiny change of the geometry of the hetero-lithic ring lasers affects the response of the instrument.
Typically mode jumps between two laser modes occur producing spikes in the output.
The presence of those spikes makes the raw data unsuitable to be stored in the open access data base used in seismology.
The analysis procedure fixes this problem: fast and simple off-line algorithms have been developed to remove them. 
In the following the apparatus of GINGERINO and the analysis procedure will be described, moreover we will introduce a qualitative comparison with Polar motion signal.

\section{The apparatus of GINGERINO}   
\subsection{The apparatus}            
GINGERINO is a 3.6 m side square ring-laser installed in the underground laboratory of Gran Sasso near L'Aquila - Italy.
It has been realised with the mechanical parts of our first prototype G-Pisa \cite{Belfi2012b}, utilizing with little modifications the geosensor scheme developed by U. Schreiber\cite{Mlinar1997}.
In order to align optically the ring cavity, the orientation of each mirror can be changed using levers driven by micrometric screws.
The under vacuum mirror box contains bearings and elastic joints for the adjustment of the pitch and yaw tilts of the mirror.
We call this model hetero-lithic, to distinguish it from G which is monolithic.
It is generally accepted that future array of ring-laser should be based on hetero-lithic structures.
ROMY, the array now operative inside the seismological observatory of Bavaria is based on a hetero-lithic model.
GINGERINO is mounted on a granite slab, which is connected to the ground by a reinforced concrete structure.
The whole structure is anchored to the ground only in the centre, and the vacuum tubes and the discharge are attached to the granite\cite{Belfi2017}.
\begin{figure}[htbp]
	\centering
	\includegraphics[width=.5\linewidth]{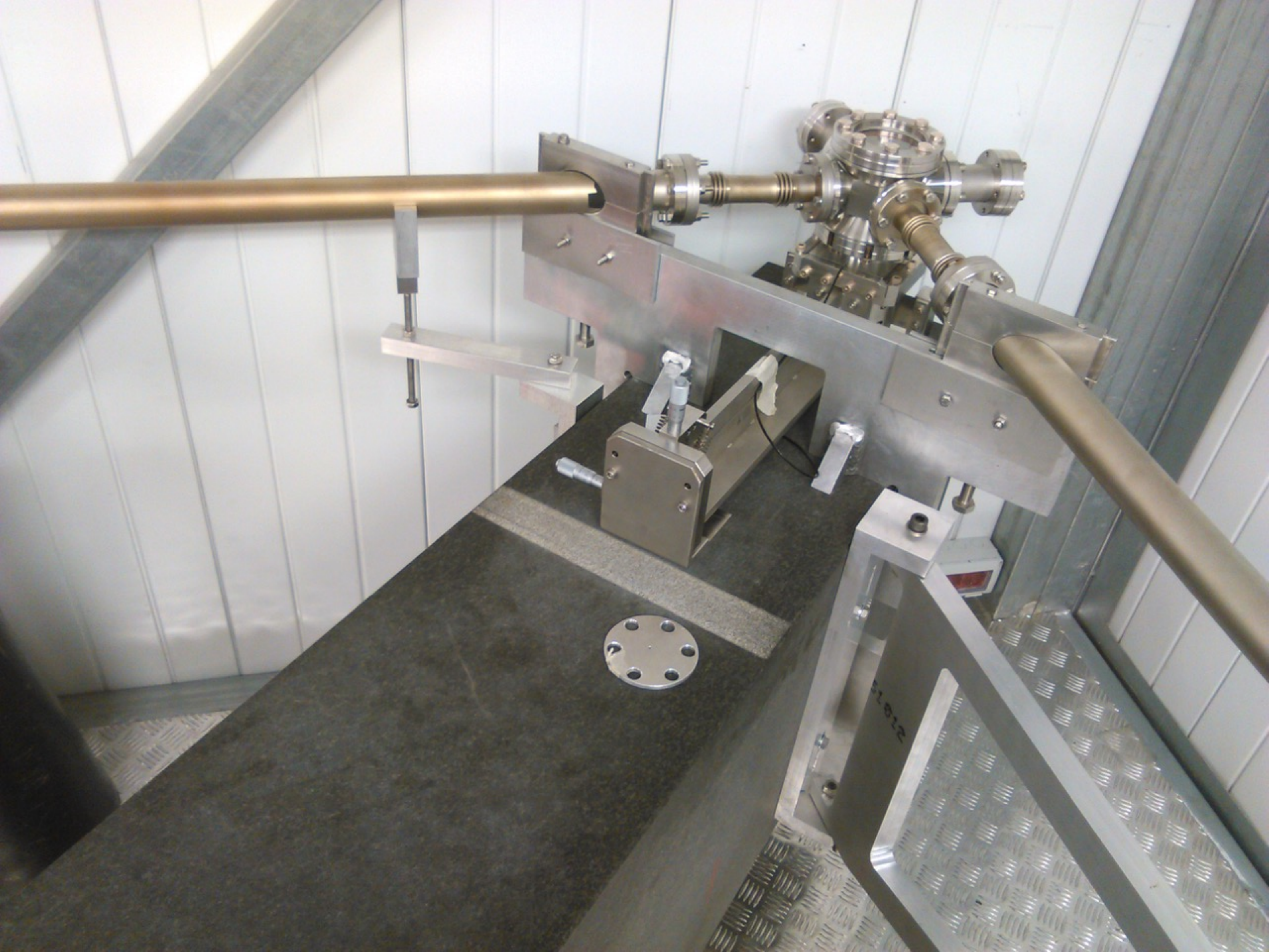} 
	\caption{Detail of one of the 4 mirror boxes, the levers to orientate the mirror, and the structures to stiffen the mechanics are visible.}
	\label{pipes}
\end{figure}
Since May 3, 2017 GINGERINO is taking data continuously.
It is free running because the geometry control has not been yet completed.\\
We implemented only a control to keep constant the optical power of one of the two  counter propagating laser beams by acting on the discharge current.
Indeed, the laser gain must be set close to threshold, in order to warrant single longitudinal mode operation.
This control is needed in order to avoid abrupt shut-down of the laser.
Our optoelectronic system acquires the following signals: the Sagnac interferometric signal, the intensities of the clockwise and counter-clockwise laser beams, ($I_{cw}$ and $I_{ccw}$), and a Discharge plasma florescence Monitor (DM), which is taken by a photodiode placed very close to the pyrex capillary.
Monobeam signals are narrow-band optically filtered around the 633 nm laser wavelength to cut plasma fluorescence.
These raw data are acquired and stored at 5kHz acquisition rate. 
Additional high sensitivity devices are co-located: one tilt meter with 1 nrad resolution (2-K High Resolution Tiltmeter HRTM, Lipmann), an high performance seismometer (Trillium 240s), a pressure meter and temperature probes.
The seismometer (owned by the Italian INGV, Istituto Nazionale di Geofisica e Vulcanologia) produces data that are stored in a data base at 200 Hz, while the other monitors (temperature, tilt and pressure) are acquired by the GINGERINO DAQ system and stored at 1 Hz.
Presently, because of a system failure, temperature data are not available.\\
Seismology is a science based on the observation of ground motions.
Two types of measures are routinely implemented, translation and strain, observed by inertial seismometer and by strain
meters based on optical or mechanical principles.
However a full description of the ground movements requires also the acquisition of a third type of measurement, namely rotations \cite{aki2002quantitative}.
The need of co-located translation and rotation observables at different sites is outlined in Ref. \cite{Igel2007}, in order to investigate the local velocity structure.
Rotational signals induced by seismic waves have a quite small amplitude.
A strong seismic wave with a linear acceleration of 1 $mm/s^2$ produces a rotation velocity amplitude of some $10^{-7}$ $rad/s$, while microseismic (around $0.1 Hz$) rotational background noise should be expected smaller than $10^{-10}$ rad/s \cite{Hadziioannou2012}.
Then, rotational seismology requires high sensitivity sensors.
Large frame ring laser gyroscopes demonstrated a unrivalled sensitivity level.
Impressive data were collected on the ring laser G located at the Geodetic Observatory in Wettzell - Germany.
Already in 2003 was reported a fully consistent observation of the rotation around a vertical axis and of the the linear translational motion recorded in by co-located seismometers \cite{Cochard2005}.
The sensitivity of G in the $0.01-0.5\ Hz$ region was pushed later at the level lower than $10^{-11}\ rad/s$, allowing a clear observation of the secondary microseismic peak \cite{Hadziioannou2012}.
Other less sophisticated laser gyroscopes proved to be capable to produce relevant information for seismology.
As an example, our first prototype G-Pisa, which was located at the VIRGO\footnote{\href{https://www.ego-gw.it}{https://www.ego-gw.it}} gravitational antenna site, was able in 2011 to record with good resolution the rotational signal around a horizontal axis induced by the $Mw = 9.0$, March 2011, Japan Earthquake \cite{Belfi2012}.\\
GINGERINO constitutes an important step for rotational seismology.
Its location, under more than $1 km $ of rock at the LNGS site, isolates effectively the apparatus from the anthropic and meteorologic noise and guarantees a high level of environmental stability.
Also the level of microseismic noise, as detected by the co-located linear seismometers, is significantly low.
During its acquisition time, started in Summer 2016, a lot of seismic events has been observed.
In particular GINGERINO acquired in almost near field the Central Italy seismic sequence, whose epicentres were in a radius of less than $100\ km$ \cite{Simonelli}, demonstrating that it is possible to identify the seismic waves back-azimuth (bottom of \figref{russia}).
In the occurrence of the Mexico MW 8.1 Earthquake the data allowed the demonstration of the G1,G2,G3,G4 onsets of the surface Love waves \cite{Simonelli2017}. \figref{russia} (top) we show the seismic event of July 17, Kamchatka.
\begin{figure}[htbp]
	\centering
	\includegraphics[width=\linewidth]{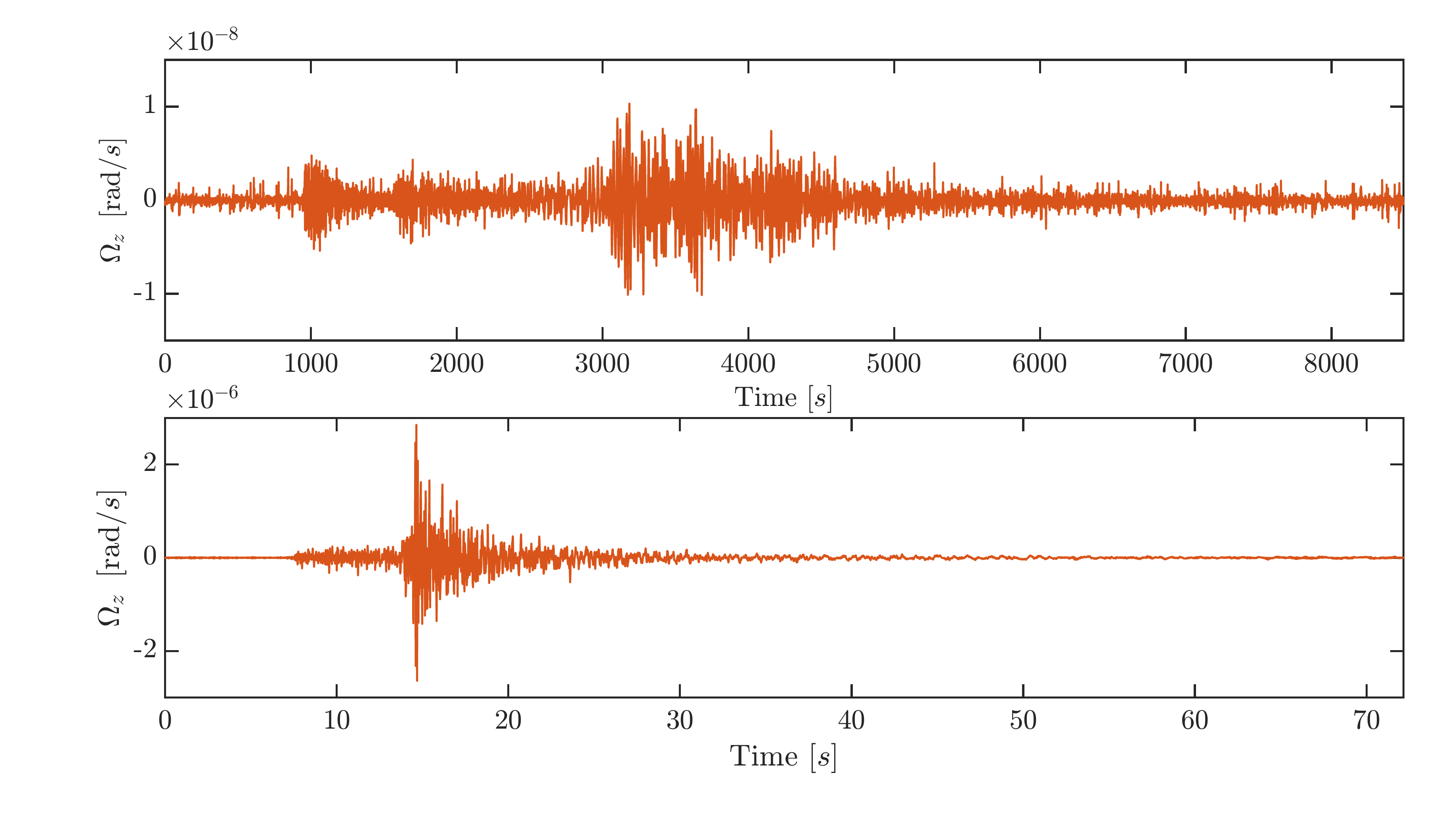}
	\caption{Typical signal from far away strong Earthquake, the tele-seismic event on July 18 2017, around  $00:00$ (UTC), Kamchatka, Russia, M 7.6 (TOP) and a typical signal of regional Earthquake, on October 16 2016 09:32:35 (UTC), Rieti, Italy Mw 4.0 (BOTTOM)  \label{russia}}
\end{figure}


\subsection{Data acquisition}
The offline analysis is composed of 2 steps.
In the first one we evaluate from the raw data the instantaneous Sagnac frequency $f_S$, and all the parameters of the laser kinematics.
The instantaneous Sagnac frequency $f_S$ is reconstructed using the Hilbert transform, decimated and stored at 10Hz.
This data set is useful to evidence any fast dynamics related to seismic regional event.
In the meantime the raw data are bandpass filtered around the nominal Sagnac frequency of 280.4 Hz with 2 Hz bandwidth, subdued to the Hilbert transform and finally stored at 1 Hz.
This data set is useful for the analysis of very low frequency features ( as solid Earth tides, and polar motion,).\\
The backscattering ($BS$) is the optical coupling of the CW with the CCW laser beam because of light scattering on the surface of each mirror.
Its temporal fluctuations produces a significative low frequency noise on the Sagnac frequency that we can estimate by analysing the mono beams signals.
Indeed, $I_{cw}$ and $I_{ccw}$, contains at the Sagnac frequency information related to the interference with the light backscattered on the mirrors. Applying the Hilbert transform to this data, we can find the phase ($\phi_{cw}$ and $\phi_{ccw}$) and the amplitude ($A_{cw}$ and $A_{ccw}$)
Thus, one estimation of the backscattering term is given by \cite{Hurst2014}:
\begin{equation}
BS = \frac{A_{cw} A_{ccw}}{I_{cw} I_{ccw}}\times\cos(\phi_{cw}-\phi_{ccw})
\end{equation}
As for the Sagnac frequency $f_S$, the quantities $BS$, and $DM$ are decimated and stored at 1 and 10 Hz.
Further, a very convenient parameter is the fringe contrast $FC$ defined as $\frac{IS_{max}-IS_{min}}{IS_{max}}$, where $IS$ is the instantaneous intensity of the Sagnac signal.
$FC$ is evaluated from the 5kHz sampled Sagnac signal and stored with the other parameters.\\
The second step can imply different procedures finalised to specific targets.
For instance, in the seismological application, the analysis goes further by implementing a fast procedure to eliminate the data features due to laser dynamics that can cause unwanted triggers for the seismology analysis.
From another side, the low frequency part of the signal  can be analysed to investigate the causes of instrumental instability.
This could suggest how to improve the low frequency response of the apparatus to investigate the presence of global signals such as Polar motion and solid Earth tides.
\subsection{The main characteristics of the ring-laser signal}
GINGERINO is a single axis ring laser gyroscope.
This means that it is sensitive only to one component of the angular rotation vector.
Its output is the Sagnac frequency $f_S$, which is connected to instantaneous rotation rate around the ring axis $\Omega=\Omega_E \cos(\theta+\delta\theta)+\delta\omega$, where $\Omega_E$ is the instantaneous Earth rotation rate, $\theta$ is the local colatitude, $\delta_\theta$ is the inclination angle in the meridian plane of the ring axis with respect to the zenith and $\delta\omega$ indicates the local rotation effects, including those coming from the apparatus itself.
Moreover, the Sagnac frequency is also affected by the laser dynamics.
This effect can be phenomenologically correlated to the value of backscattering, the intensities difference of the two counter propagating beams and to the plasma fluorescence intensity, estimated by the Discharge Monitor $DM$.
Assuming that those terms are small it is possible to expand in series the response of the ring-laser in function of the $BS$, $\Delta I=I_c-I_{cc}$ and $DM$ \cite{Hurst2014}
\begin{equation}
 f_S = 4(S+\delta S)( \Omega_E +\delta\omega)\cos{(\theta +\delta\theta)}+ a_1\times BS + a_2\times \Delta I + a_3\times DM
\label{fondamentale}
\end{equation}
where $S$ an $\delta S$ are the geometrical scale factor and its variation, $\theta$ is the local  colatitude ($47.55^o$ for LNGS) and $\delta\theta$ is its variation.
By fitting the parameters $a_i$ we can isolate and subtract the contributes of $BS$, $\Delta I$, $DM$ from $f_S$ in \eqref{fondamentale}.
From now on what it remains will be called "residuals".
At this point, it is evident that it is not possible to discriminate among the effect of $\delta S$, $\delta\omega$ and $\delta\theta$ using a single gyroscope.
So any variation of $f_S$ can be induced by a change of scale factor ($\delta S$), angular velocity ($\delta\omega$) or inclination angle to the north ($\delta\theta$).
These residuals from \eqref{fondamentale}, expressed as variations of the Sagnac frequency $\delta f_S$, contain the relevant informations.
It is useful to define the following quantities: $k_S$, $k_\omega$ and $k_\theta$, with value for GINGERINO:
\begin{eqnarray}
\delta S = k_S \delta f_S, \quad k_S=\frac{1}{\Omega_E\cos{\theta}}, \quad k_S=1.98\times10^4\quad s\\
\delta\omega = k_\omega \delta f_S, \quad k_\omega=\frac{1}{S\cos{\theta}},\quad k_\omega = 1.76\times 10^{-7}\quad rad/s/Hz\\
\delta\theta = k_\theta \delta f_S, \quad k_\theta = \frac{1}{S\Omega_E \sin{\theta}},\quad k_\theta = 1/306 \quad rad/Hz
\end{eqnarray}
These expressions relate a Sagnac frequency change to $\delta S$, $\delta\omega$ and $\delta\theta$.
In general the variations $\delta S$ can be reduced controlling the geometry of the square cavity.
At present GINGERINO is free running, and the variations of $S$ must be taken into account in the data analysis.
\subsection{Main characteristics of the data set}
\begin{figure}[htbp]
	\centering
	\includegraphics[width=\linewidth]{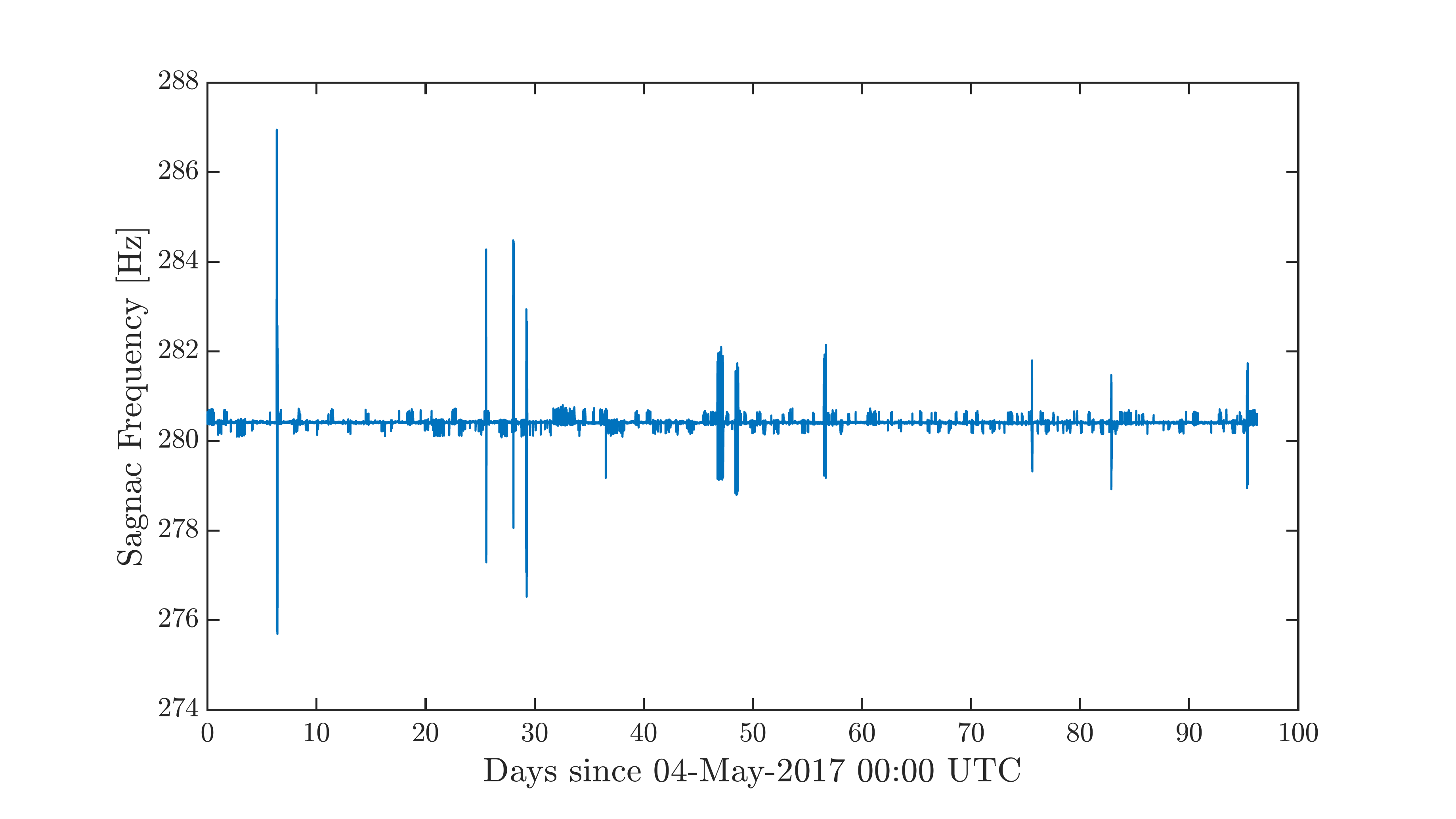} 
	\caption{97 days of data, raw data.}
	\label{fig1}
\end{figure}

\begin{figure}[htbp]
	\centering
	\includegraphics[width=\linewidth]{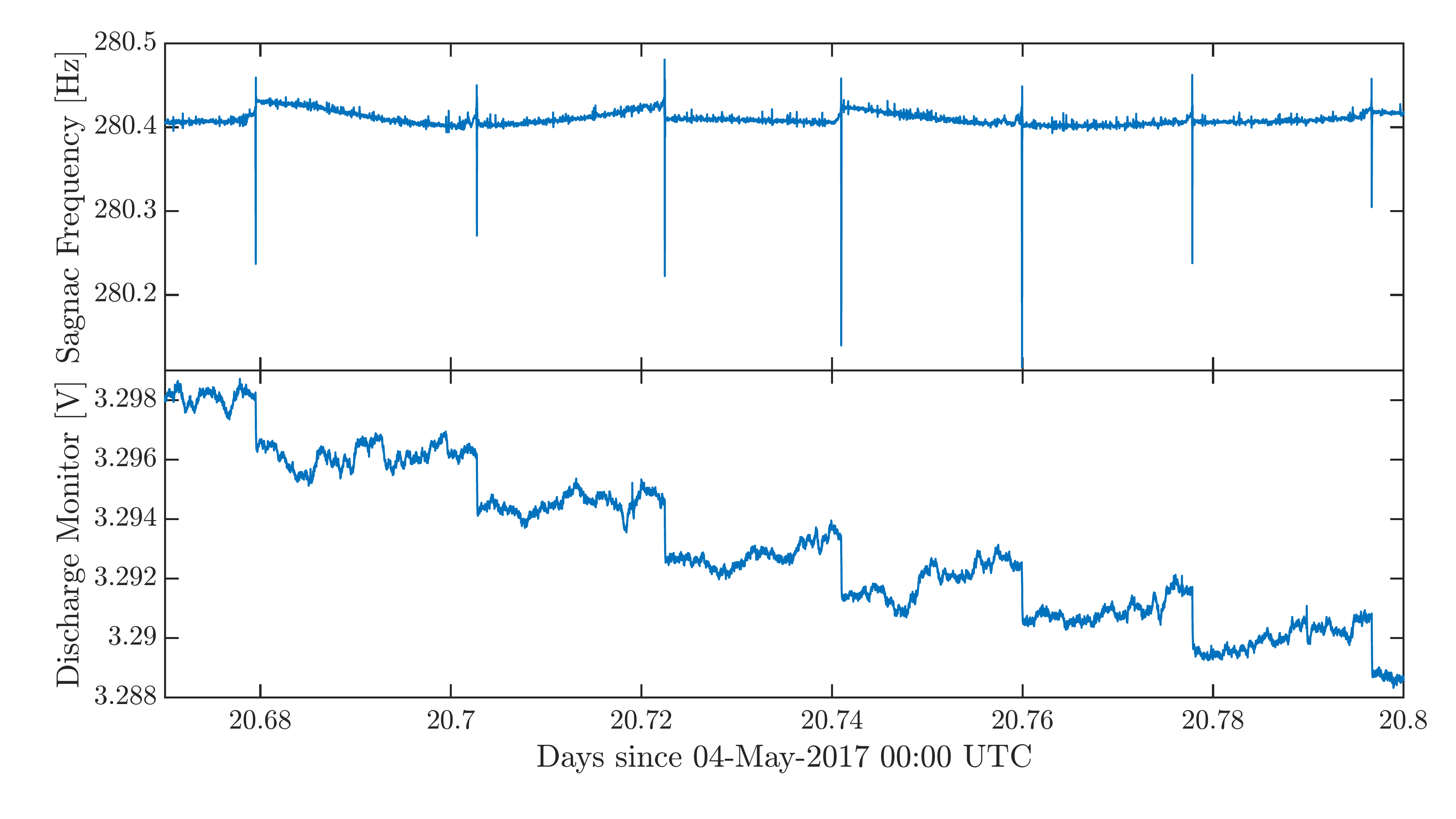} 
	\caption{Mode jumps observed in the Sagnac frequency and the discharge monitor.}
	\label{fig2}
\end{figure}

\begin{figure}[htbp]
	\centering
	\includegraphics[width=\linewidth]{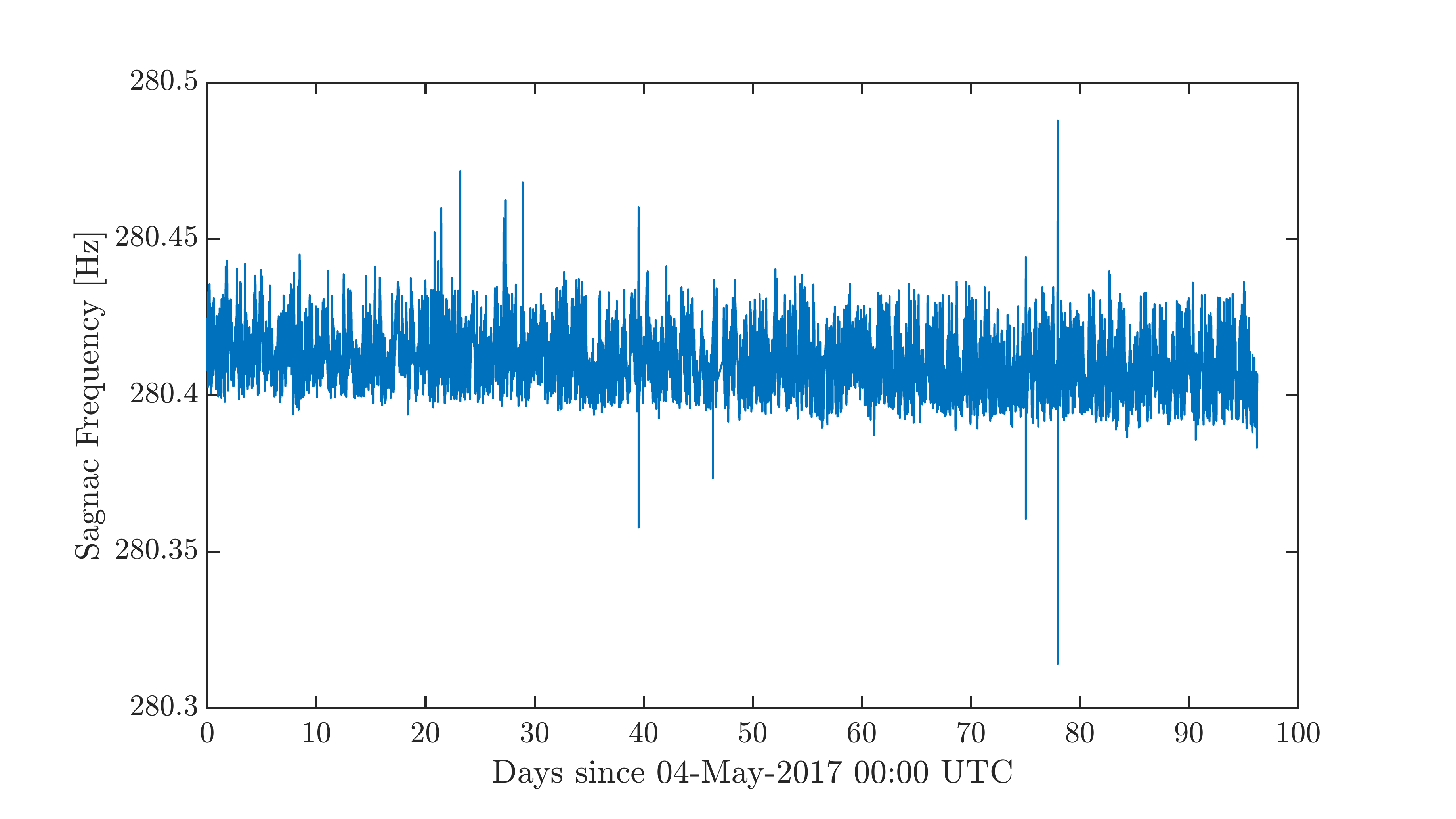}
	\caption{Sagnac frequency without split mode and mode jumps. Two Earthquakes are well visible: the M6.4 of June 12 (Greece) and the M6.7  of July 21  (Turkey).}
	\label{fig4}
\end{figure}

\begin{figure}
	\centering
	\includegraphics[width=\linewidth]{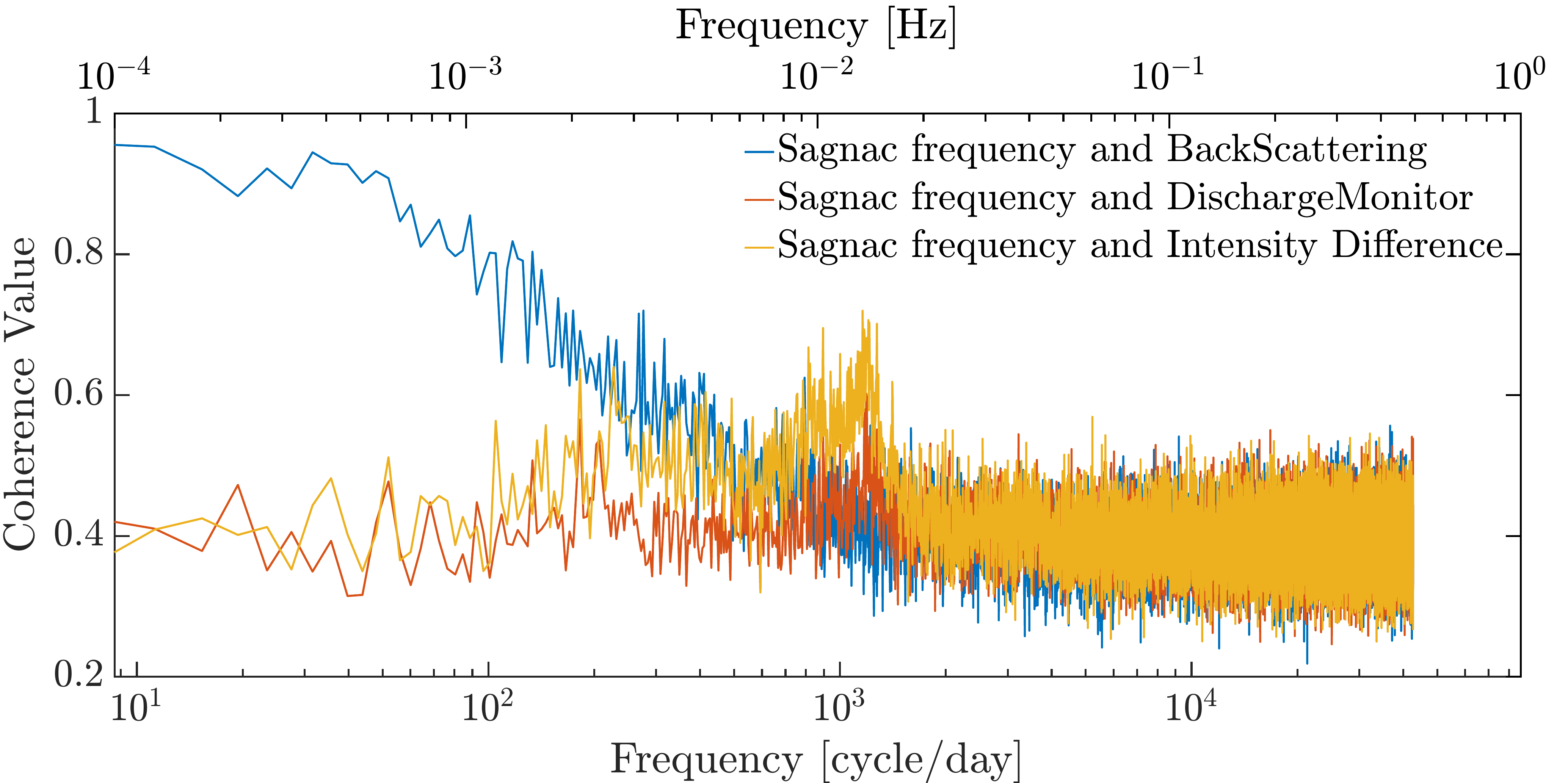}
	\caption{Square root of the magnitude-squared coherence of the Sagnac frequency $f_S$ with the backscatter $BS$, difference of intensities $\Delta I$ and discharge monitor $DM$. The calculated coherences have been filtered and decimated by a factor of 100}
	\label{fig5}
\end{figure}

\begin{figure}
	\centering
	\includegraphics[width=\linewidth]{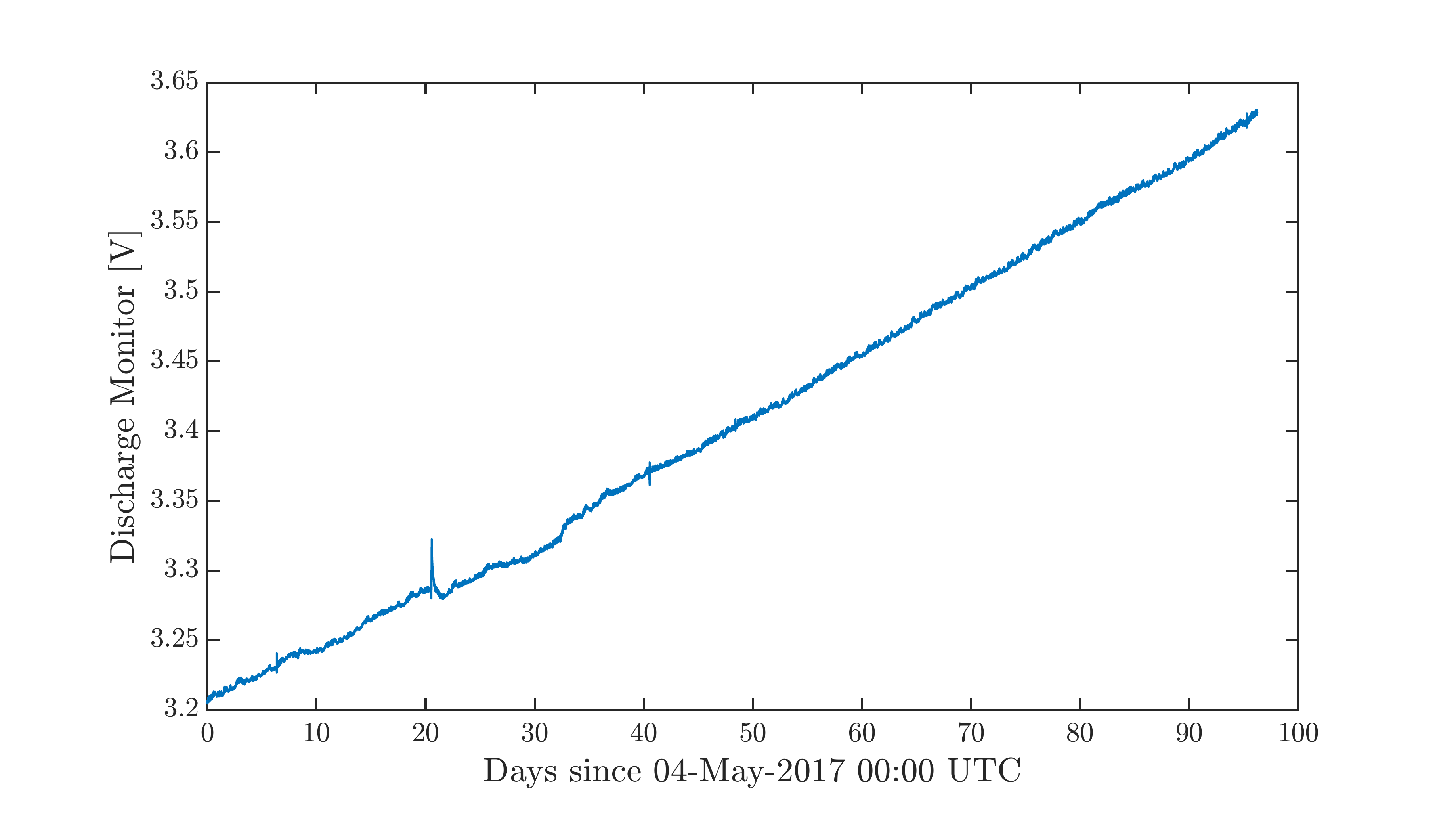}
	\caption{Discharge Monitor during the whole period.}
	\label{DM2}
\end{figure}

The whole set of data are shown in \figref{fig1}.
Well visible are the 10 portions of data with GINGERINO in split-mode, i.e. the two counter-propagating laser beams are oscillating on different longitudinal modes, with a frequency difference equal to the free spectral range of the cavity (around $21$ MHz).
GINGERINO operated in split mode for about $1.5\%$ of the time, the longest split mode was $\sim 13$ hours, and the average about 4 hours.
Looking deeper it is possible to see also the mode jumps (points where the two counter-propagating beams change mode almost at the same time).
In \figref{fig2} mode jumps are well visible in the Sagnac signal and in the plasma monitor.
In the 90 days of records we observe more than 1000 mode jumps, and it is straightforward to see that they happen very rapidly, affecting only a few seconds.
We have built a software routine which automatically identifies mode jumps and split mode. It utilizes the  Fringe Contrast (FC) and two kind of threshold levels: the lower one to recognize the split mode periods and the higher one to recognize the mode jumps.
In this way, the bad portions of the data, are identified and then eliminated and replaced by linear interpolation.
It is in principle possible to replace the data eliminated with smarter procedure, in order not to miss information but we have chosen the linear interpolation since is the simplest one.
In \figref{fig4} we show the Sagnac frequency cleaned through these procedure.
As previously said, the laser dynamics is affected by backscattering, by the difference of the two modes intensities and by the the plasma processes estimated through the discharge monitor $DM$.
In \figref{fig5} we show the coherence of the Sagnac frequency $f_S$ with respectively $BS$, $\Delta I$ and $DM$.
The working point of laser is set close to the threshold, in order to avoid multimode operation.
The output power of one of the two counter propagating beams is opto-electronically controlled by regulating the voltage of the electronic discharge.
Since the gain curve of the gas is not flat and GINGERINO is free running, when the perimeter of the cavity and the wavelength change, the control modifies the voltage in order to keep constant the output power.
In short, $DM$ gives information of the long period perimeter modifications \figref{DM2}.
All the three probes ($BS$, $\Delta I$ and $DM$) are used during the analysis process in order to clean up the output signal and to study the low frequency behaviour of the gyroscope with the residuals $\delta f_S$.   
The first step is to remove the very low frequency trend using $DM$.
This is done by subtracting the linear contribution by using a least square method.
In the second step, each portion of data between two split modes are analysed separately and the contributions from backscattering and intensity difference are subtracted by using a linear regression.  
\begin{figure}[htbp]
	\centering
	\includegraphics[width=\linewidth]{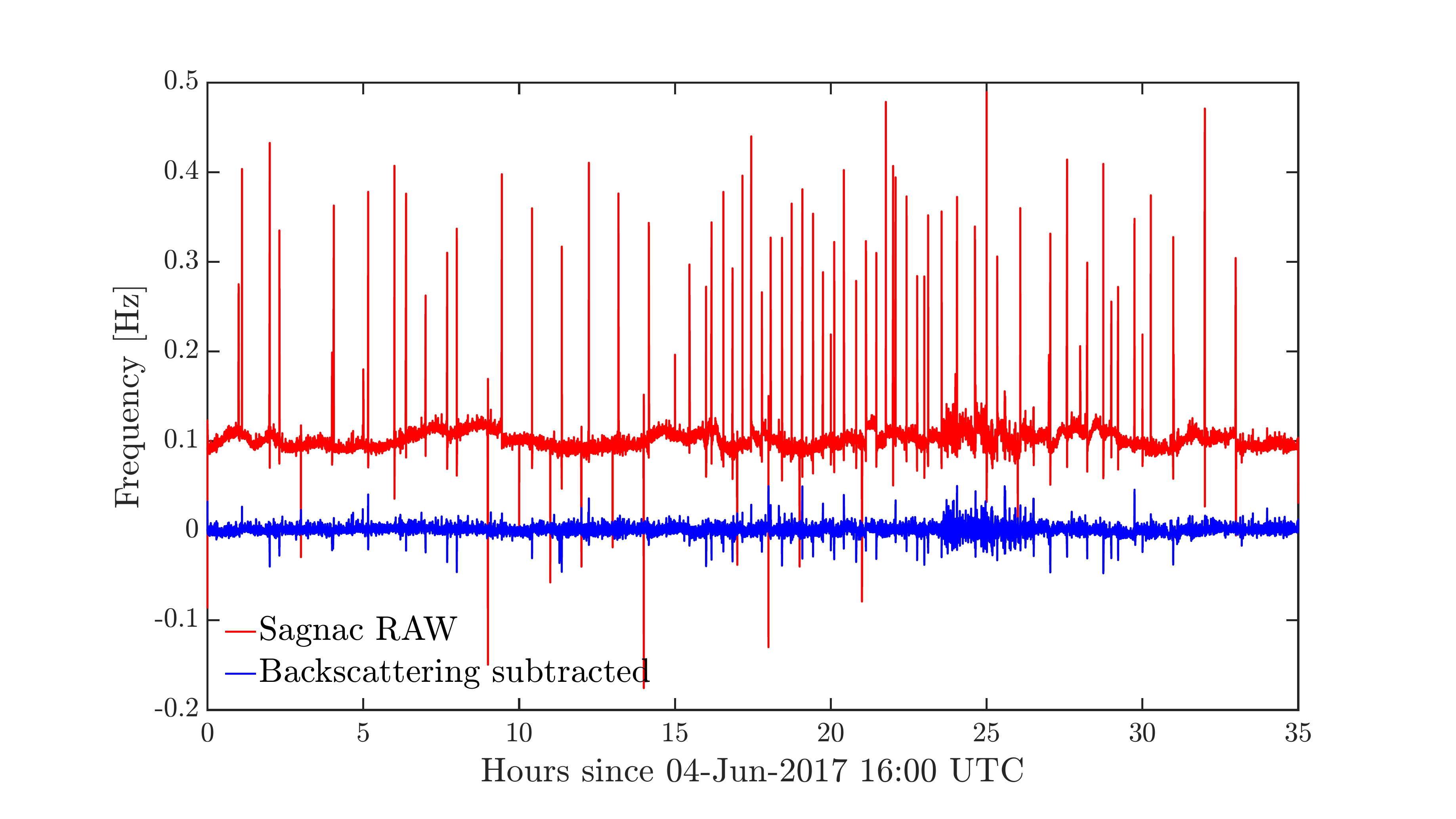}  
	\caption{The residual after backscattering subtraction is compared with the raw data, in both curves the average is subtracted and the raw data are shifted up by 0.1Hz.}
	\label{fig3}
\end{figure}
The analysis is focused on the long time behaviour, with particular attention to the response at frequency around $1$ cycle/day, to see whether the signal induced by polar motion and solid Earth tide are visible.
We have investigated several procedure, and it has been immediately evident that the subtraction was more effective decimating the data. 
The effectiveness of this procedure is evident comparing the raw data with the residuals (\figref{fig3}).
\figref{freq_rad} shows the  residuals expressed as angular rotation on the left and change of the relative angle on the right one.
\begin{figure}[htbp]
	\centering
	\includegraphics[width=\linewidth]{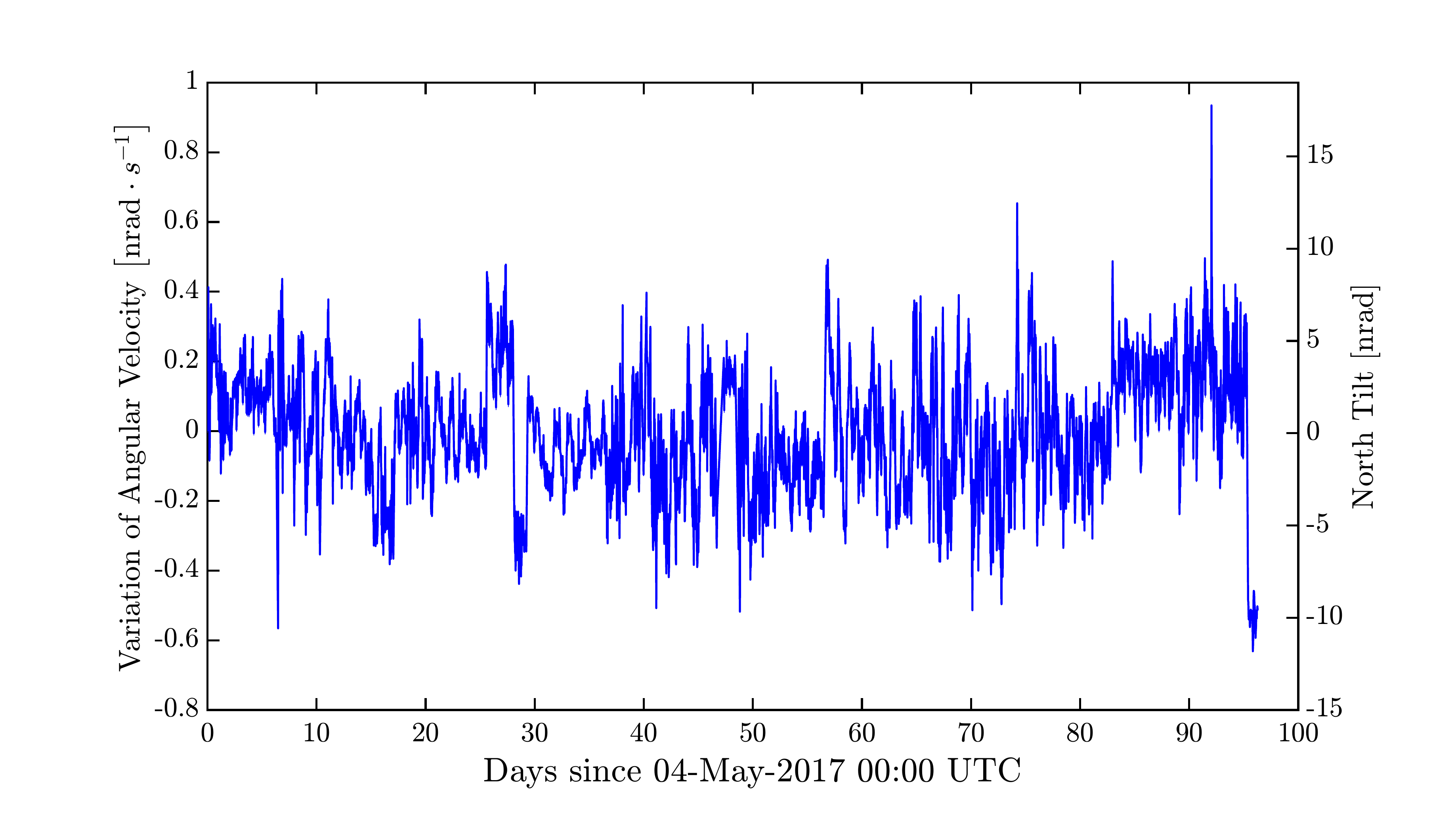}
	\caption{Residuals expressed as angular velocity (left) or as tilt angle with respect to the north south direction (right). GINGERINO is a single axis gyroscope and the residuals can be interpreted as angular velocity or change in the relative angle with the rotation axis.}
	\label{freq_rad}
\end{figure}

\subsection{Comparison with expected Polar Motion and solid Earth tide signals}
\begin{figure}[htbp]
	\centering
	\includegraphics[width=\linewidth]{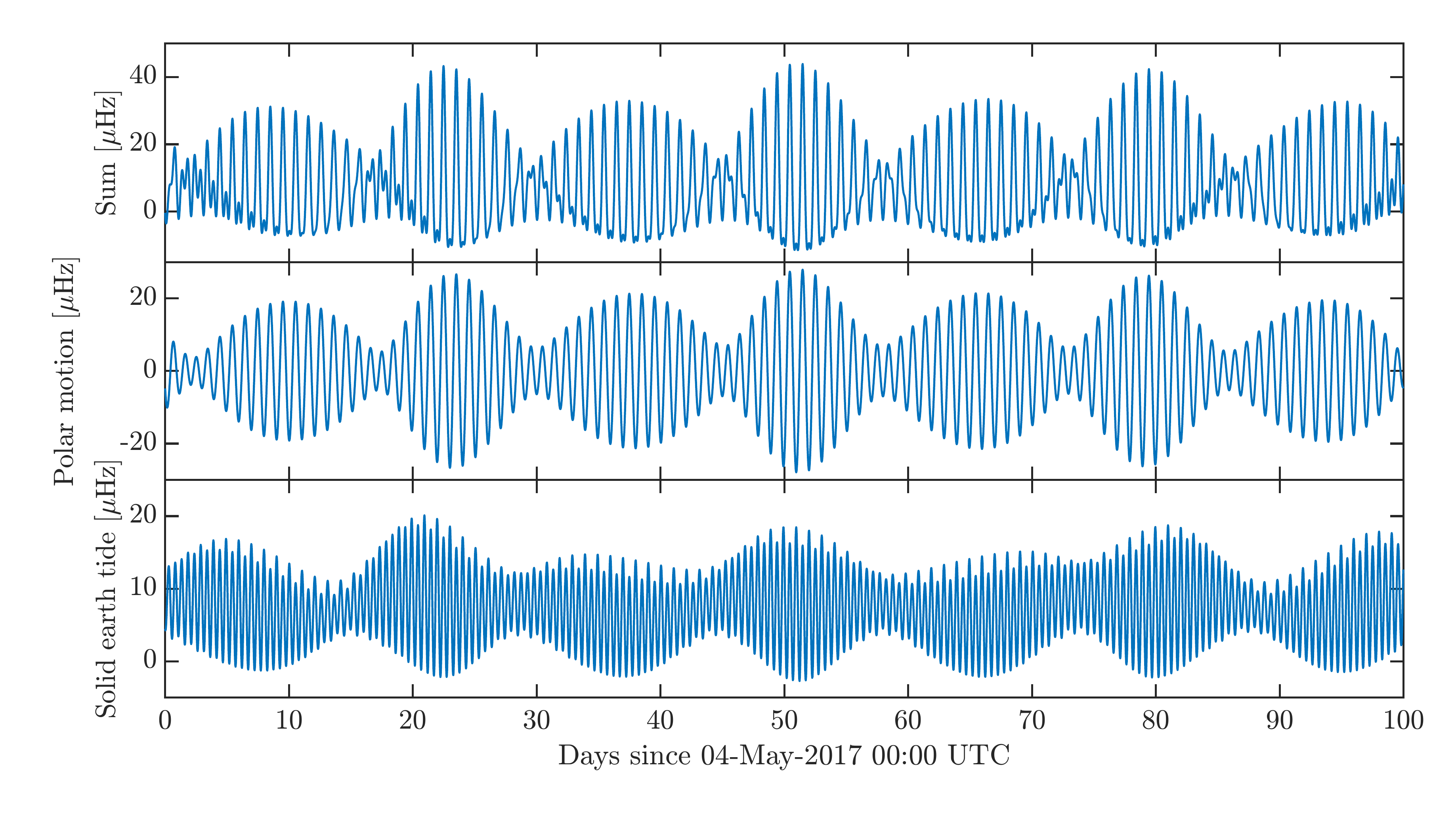}
	\caption{Synthetic effect on GINGERINO. TOP: summed polar motion and Earth tide, MIDDLE polar motion only, BOTTOM solid Earth tide.}
	\label{pmTide}
\end{figure}

\begin{figure}[htbp]
	\centering
	\includegraphics[width=\linewidth]{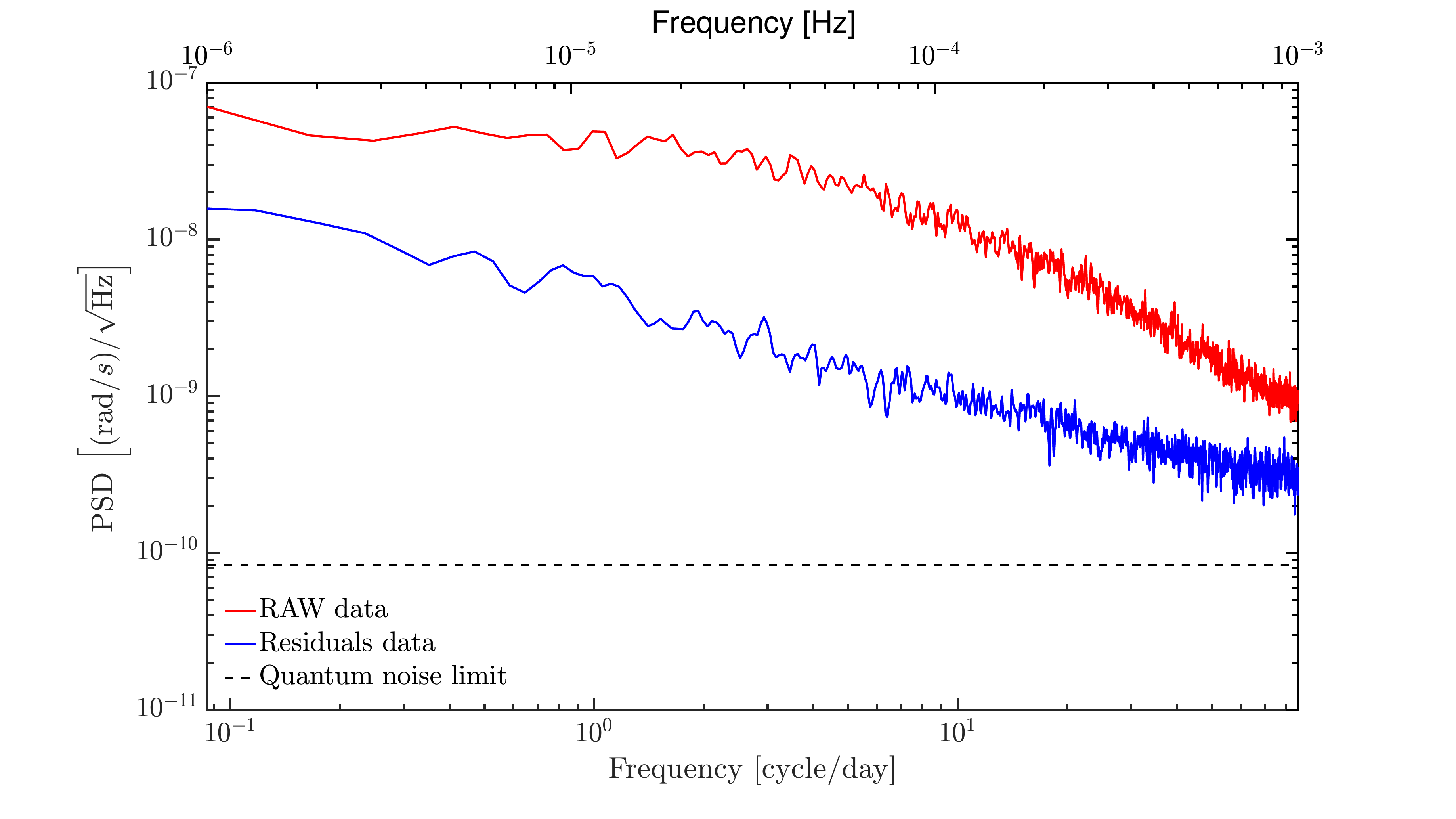}
	\caption{Comparison of the power spectral density of the residuals, the RAW data set and the quantum noise limit.}
	\label{fig8}
\end{figure}

\begin{figure}[htbp]
	\centering
	\includegraphics[width=\linewidth]{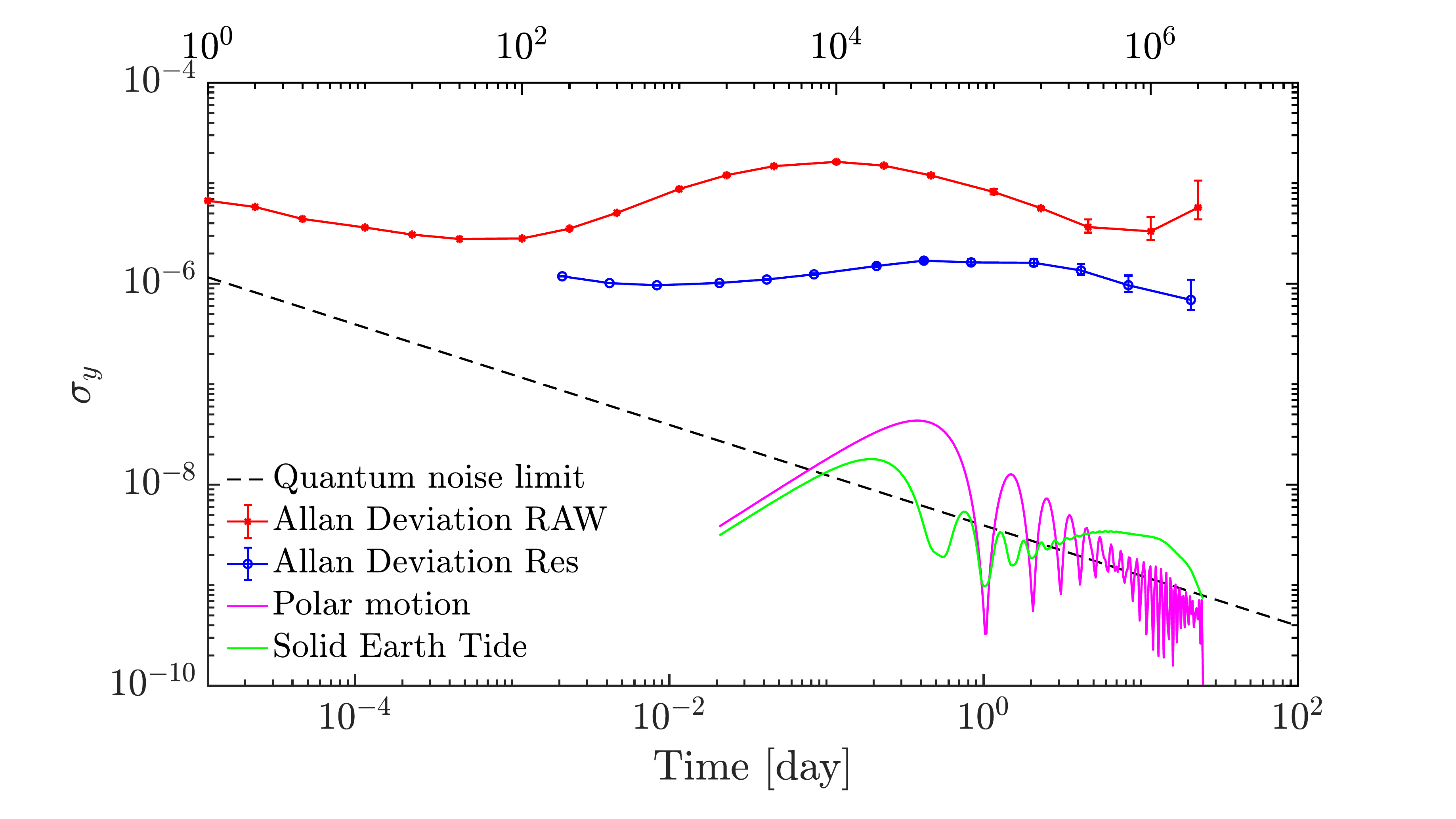}
	\caption{Overlapping Allan deviation of raw data (red curve) and the residuals (blue curve) compared with the quantum noise limit (dashed line) and the Polar motion (magenta curve) and Solid Earth Tide (green curve) of GINGERINO site. Due to computational needs the residuals overlapping Allan deviation has been calculated using data sampled at $180s$}
	\label{Allan}
\end{figure}

A horizontal gyroscope as GINGERINO is affected by the relative motion of the moon and the sun.
Among all possible effects, the dominating ones are the polar motion and the north-south tilt caused by the solid Earth tide.
The polar motion is basically a signal around 1 cycle/day, while the Earth tide has several components.
The used model is based on the simulation developed for G Wettzell by Andre' Gebauer \cite{privGebauer} (\figref{pmTide}).
To have an idea of the performances of our gyroscopes we compare in \figref{fig8} the power spectral density of the RAW data, residuals one and the quantum shot noise of our instrument.
Another important way to show the performances of our instrument is the comparison between the overlapping Allan deviation of the Sagnac raw data and the residuals one, the overlapping Allan deviation of the polar motion and solid Earth tide and the quantum shot noise of our instrument \figref{Allan}.
This value is estimated by \cite{Schreiber2008}:
\begin{equation}
\Omega_N=\frac{cP}{4QA}\sqrt{\frac{h\nu}{P_0t}}
\end{equation}
where $c$ is the speed of light, $P$ and $A$ are the ring perimeter and its area, $Q$ the cavity quality factor, $h$ the Plank constant, $\nu$ the laser frequency, $P_0$ the cavity power losses and $t$ the integration time.
For GINGERINO $\Omega_N\simeq8.43\times10^{-11}\sqrt{t}\ [\mathrm{rad}/s]$ ($1.16\times10^{-6}\sqrt{t}$ in relative units)\footnote{these values are evaluated considering seconds as units for $t$}%
\section{Discussion and Conclusions}
GINGERINO is the first high sensitivity ring laser based on hetero-lithic structure, which operated unattended for several months with a duty cycle higher than $95\%$ and a sensitivity around $0.1$ nrad/s for 1 second integration time.
We analysed the first 96 days of continuous data throw suitable software routines in order to make them available to a broader community.
A large number of disturbances caused by mode jumps are present in the data.
It is necessary to eliminate these noise sources to avoid fake triggers.
A fast procedure, suitable to be implemented on-line, has been developed at this purpose.\\
The low frequency part of the spectrum is of primary interest, but below $1$ mHz the systematics of the laser, as backscatter noise, dominates.
Using the procedure described in this paper the gyroscope Allan variance has been improved of about one order of magnitude, reducing the noise coming from backscattering and other systematics of the laser.\\
The data exhibits a large number of mode jumps, the developed procedure is not able to completely remove the disturbances induced by these ones, so part of the excess noise could be due to them.
The number of mode jumps are rather high and not consistent with the typical temperature variation inside the GINGERINO site.
One possible explanation is that the laser cavity is not stable enough in size and orientation, further study is necessary to explain the high occurrence of mode jumps, for that the probes of temperature are necessary.
GINGERINO mechanics is $10$ years old.
In the meantime, more stable mechanical solution have been developed.
ROMY, for example, has a more rigid structure. GP2, our new prototype, which has two stages of alignment, rough and fine.
In the near future  we plan to fix the DAQ problems and the temperature probes.\\
The mirrors losses play a crucial role in this kind of apparatus, using mirrors with lower losses the shot noise limit improve, and the noise coming from the backscattering, which is the dominant noise source, decreases.
At present the losses of the installed mirrors are more than a factor $10$ higher than the ones installed in G. A new set of mirrors will be installed in GINGERINO before summer 2018, tests are in progress in order to test the quality of these new mirrors.
Nowadays the system is still running and the analysis will continue in order to further improve the comprehension of the very low frequency response. 

\section*{ } 

\newpage
%


\bibliographystyle{plain}

\begin{thebibliography}{10}
	
	\bibitem{aki2002quantitative}
	K~Aki and P~G Richards.
	\newblock {\em {Quantitative Seismology}}.
	\newblock Geology (University Science Books).: Seismology. University Science
	Books, 2002.
	
	\bibitem{Belfi2012}
	J.~Belfi, N.~Beverini, F.~Bosi, G.~Carelli, A.~{Di Virgilio}, E.~Maccioni,
	A.~Ortolan, and F.~Stefani.
	\newblock {A 1.82 m 2 ring laser gyroscope for nano-rotational motion sensing}.
	\newblock {\em Applied Physics B: Lasers and Optics}, 106(2):271--281, 2012.
	
	\bibitem{Belfi2017}
	Jacopo Belfi, Nicol{\`{o}} Beverini, Filippo Bosi, Giorgio Carelli, Davide
	Cuccato, Gaetano {De Luca}, Angela {Di Virgilio}, Andr{\'{e}} Gebauer, Enrico
	Maccioni, Antonello Ortolan, Alberto Porzio, Gilberto Saccorotti, Andreino
	Simonelli, and Giuseppe Terreni.
	\newblock {Deep underground rotation measurements: GINGERino ring laser
		gyroscope in Gran Sasso}.
	\newblock {\em Review of Scientific Instruments}, 88(3):034502, mar 2017.
	
	\bibitem{Belfi2012b}
	Jacopo Belfi, Nicol{\`{o}} Beverini, Giorgio Carelli, Angela {Di Virgilio},
	Enrico Maccioni, Gilberto Saccorotti, Fabio Stefani, and Alexander
	Velikoseltsev.
	\newblock {Horizontal rotation signals detected by “G-Pisa” ring laser for
		the M w = 9.0, March 2011, Japan earthquake}.
	\newblock {\em Journal of Seismology}, 16(4):767--776, oct 2012.
	
	\bibitem{Bosi2011}
	F.~Bosi, G.~Cella, A.~{Di Virgilio}, A.~Ortolan, A.~Porzio, S.~Solimeno,
	M.~Cerdonio, J.~P. Zendri, M.~Allegrini, J.~Belfi, N.~Beverini, B.~Bouhadef,
	G.~Carelli, I.~Ferrante, E.~Maccioni, R.~Passaquieti, F.~Stefani, M.~L.
	Ruggiero, A.~Tartaglia, K.~U. Schreiber, A.~Gebauer, and J-P.~R. Wells.
	\newblock {Measuring gravitomagnetic effects by a multi-ring-laser gyroscope}.
	\newblock {\em Physical Review D}, 84(12):122002, dec 2011.
	
	\bibitem{Cochard2005}
	Alain Cochard, H.~Igel, B.~Schuberth, W.~Suryanto, A.~Velikoseltsev,
	U.~Schreiber, J.~Wassermann, F.~Scherbaum, and D.~Vollmer.
	\newblock {Rotational Motions in Seismology: Theory, Observation, Simulation}.
	\newblock In {\em Earthquake Source Asymmetry, Structural Media and Rotation
		Effects}, pages 391--411. Springer-Verlag, Berlin/Heidelberg, 2005.
	
	\bibitem{DiVirgilio2017}
	Angela D.~V. {Di Virgilio}, Jacopo Belfi, Wei-Tou Ni, Nicolo Beverini, Giorgio
	Carelli, Enrico Maccioni, and Alberto Porzio.
	\newblock {GINGER: A feasibility study}.
	\newblock {\em The European Physical Journal Plus}, 132(4):157, apr 2017.
	
	\bibitem{privGebauer}
	Andr{\'{e}} Gebauer.
	\newblock {Private Communication}.
	
	\bibitem{Hadziioannou2012}
	Celine Hadziioannou, Peter Gaebler, Karl~Ulrich Schreiber, Joachim Wassermann,
	and Heiner Igel.
	\newblock {Examining ambient noise using colocated measurements of rotational
		and translational motion}.
	\newblock {\em Journal of Seismology}, 16(4):787--796, oct 2012.
	
	\bibitem{Hurst2014}
	R.B. Hurst, N.~Rabeendran, K.U. Schreiber, and J.-P.R. Wells.
	\newblock {Correction of backscatter-induced systematic errors in ring laser
		gyroscopes}.
	\newblock {\em Applied Optics}, 53(31), 2014.
	
	\bibitem{Igel2007}
	Heiner Igel, Alain Cochard, Joachim Wassermann, Asher Flaws, Ulrich Schreiber,
	Alex Velikoseltsev, and Nguyen {Pham Dinh}.
	\newblock {Broad-band observations of earthquake-induced rotational ground
		motions}.
	\newblock {\em Geophysical Journal International}, 168(1):182--196, jan 2007.
	
	\bibitem{Sagnac1914}
	G.~Sagnac.
	\newblock {Effet tourbillonnaire optique. La circulation de l'{\'{e}}ther
		lumineux dans un interf{\'{e}}rographe tournant}.
	\newblock {\em Journal de Physique Th{\'{e}}orique et Appliqu{\'{e}}e},
	4(1):177--195, 1914.
	
	\bibitem{Schreiber2008}
	K.~U. Schreiber, J.~P~R Wells, and G.~E. Stedman.
	\newblock {Noise processes in large ring lasers}.
	\newblock {\em General Relativity and Gravitation}, 40(5):935--943, 2008.
	
	\bibitem{privSchreiber}
	Karl~Ulrich Schreiber.
	\newblock {Private Communication}.
	
	\bibitem{Schreiber2011}
	Karl~Ulrich Schreiber, T.~Kl{\"{u}}gel, J.-P.~R. Wells, R.~B. Hurst, and
	A.~Gebauer.
	\newblock How to detect the chandler and the annual wobble of the earth with a
	large ring laser gyroscope.
	\newblock {\em Physical Review Letters}, 107(17):173904, oct 2011.
	
	\bibitem{Schreiber2013}
	Karl~Ulrich Schreiber and Jon-Paul. Wells.
	\newblock {Invited Review Article: Large ring lasers for rotation sensing}.
	\newblock {\em Review of Scientific Instruments}, 84(4):041101, apr 2013.
	
	\bibitem{Simonelli2017}
	Andreino Simonelli, Jacopo Belfi, Nicol{\`{o}} Beverini, Angela D.~V. {Di
		Virgilio}, Umberto Giacomelli, Gaetano {De Luca}, and Heiner Igel.
	\newblock {Love waves trains observed after the MW 8.1 Tehuantepec earthquake
		by an underground ring laser gyroscope}.
	\newblock In {\em AGU Fall Meeting 2017, S33G-2954}, New Orleans, 2017.
	
	\bibitem{Simonelli}
	Andreino Simonelli, Heiner Igel, Joachim Wassermann, Jacopo Belfi, Enrico
	Maccioni, Angela D.~V. {Di Virgilio}, Nicol{\`{o}} Beverini, Gaetano {De
		Luca}, and Gilberto Saccorotti.
	\newblock {Rotational motions from the 2016, Central Italy seismic sequence, as
		observed by an underground ring laser gyroscope}.
	\newblock {\em Submitted at Geophysical Journal International}.
	
	\bibitem{Mlinar1997}
	G~E Stedman.
	\newblock {Ring-laser tests of fundamental physics and geophysics}.
	\newblock {\em Reports on Progress in Physics}, 60(6):615--688, jun 1997.
	
	\bibitem{Tartaglia2017}
	Angelo Tartaglia, Angela {Di Virgilio}, Jacopo Belfi, Nicol{\`{o}} Beverini,
	and Matteo~Luca Ruggiero.
	\newblock {Testing general relativity by means of ring lasers}.
	\newblock {\em The European Physical Journal Plus}, 132(2):73, feb 2017.
	
	\bibitem{Tercjak2017}
	Monika Tercjak and Aleksander Brzezi{\'{n}}ski.
	\newblock {On the Influence of Known Diurnal and Subdiurnal Signals in Polar
		Motion and UT1 on Ring Laser Gyroscope Observations}.
	\newblock {\em Pure and Applied Geophysics}, 174(7):2719--2731, 2017.
	
\end{thebibliography}

\end{document}